\documentclass[superscriptaddress, aps, prc, 10pt, amsmath, amssymb, bibnotes,
altaffilletter, twocolumn, floatfix]{revtex4-2}

\usepackage{hyperref}
\usepackage[T1]{fontenc}
\usepackage{lineno}
\usepackage{xcolor}
\usepackage{graphicx} 
\graphicspath{{/}}
\usepackage{dcolumn} 
\usepackage{bm} 
\hypersetup{colorlinks=true, citecolor=blue, urlcolor=blue, linkcolor=blue}
\usepackage{orcidlink}

\def\MJ{{\sc Majorana}}
\def\DEM{{\sc Demonstrator}}
\def\MJD{{\sc Majorana Demonstrator}}
\def\nonubb{$\beta\beta(0\nu)$}
\def\enrge{${}^{\mathrm{enr}}$Ge}

\begin{document}

\title{Search for charge non-conservation and Pauli exclusion principle violation with the {\sc Majorana Demonstrator}}

\newcommand{\ITEP}{National Research Center ``Kurchatov Institute'' Institute for Theoretical and Experimental Physics, Moscow, 117218 Russia}
\newcommand{\JINR}{Joint Institute for Nuclear Research, Dubna, 141980 Russia} 
\newcommand{\lbnl}{Nuclear Science Division, Lawrence Berkeley National Laboratory, Berkeley, CA 94720, USA}
\newcommand{\lbnle}{Engineering Division, Lawrence Berkeley National Laboratory, Berkeley, CA 94720, USA}
\newcommand{\lanl}{Los Alamos National Laboratory, Los Alamos, NM 87545, USA}
\newcommand{\queens}{Department of Physics, Engineering Physics and Astronomy, Queen's University, Kingston, ON K7L 3N6, Canada}
\newcommand{\uw}{Center for Experimental Nuclear Physics and Astrophysics, and Department of Physics, University of Washington, Seattle, WA 98195, USA}
\newcommand{\unc}{Department of Physics and Astronomy, University of North Carolina, Chapel Hill, NC 27514, USA}
\newcommand{\duke}{Department of Physics, Duke University, Durham, NC 27708, USA}
\newcommand{\ncsu}{Department of Physics, North Carolina State University, Raleigh, NC 27695, USA}	
\newcommand{\ornl}{Oak Ridge National Laboratory, Oak Ridge, TN 37830, USA}
\newcommand{\ou}{Research Center for Nuclear Physics, Osaka University, Ibaraki, Osaka 567-0047, Japan}
\newcommand{\pnnl}{Pacific Northwest National Laboratory, Richland, WA 99354, USA}
\newcommand{\ttu}{Tennessee Tech University, Cookeville, TN 38505, USA}
\newcommand{\sdsmt}{South Dakota Mines, Rapid City, SD 57701, USA}
\newcommand{\usc}{Department of Physics and Astronomy, University of South Carolina, Columbia, SC 29208, USA}
\newcommand{\usd}{Department of Physics, University of South Dakota, Vermillion, SD 57069, USA}  
\newcommand{\ut}{Department of Physics and Astronomy, University of Tennessee, Knoxville, TN 37916, USA}
\newcommand{\tunl}{Triangle Universities Nuclear Laboratory, Durham, NC 27708, USA}
\newcommand{\mpi}{Max-Planck-Institut f\"{u}r Physik, M\"{u}nchen, 80805, Germany}
\newcommand{\tum}{Physik Department and Excellence Cluster Universe, Technische Universit\"{a}t, M\"{u}nchen, 85748 Germany}
\newcommand{\williams}{Physics Department, Williams College, Williamstown, MA 01267, USA}
\newcommand{\ciemat}{Centro de Investigaciones Energ\'{e}ticas, Medioambientales y Tecnol\'{o}gicas, CIEMAT 28040, Madrid, Spain}
\newcommand{\iu}{Department of Physics, Indiana University, Bloomington, IN 47405, USA}
\newcommand{\iuceem}{IU Center for Exploration of Energy and Matter, Bloomington, IN 47408, USA}

\author{I.J.~Arnquist}\affiliation{\pnnl} 
\author{F.T.~Avignone~III}\affiliation{\usc}\affiliation{\ornl}
\author{A.S.~Barabash}\affiliation{\ITEP}
\author{C.J.~Barton}\affiliation{\usd}
\author{K.H.~Bhimani}\affiliation{\unc}\affiliation{\tunl} 
\author{E.~Blalock}\affiliation{\ncsu}\affiliation{\tunl} 
\author{B.~Bos}\affiliation{\unc}\affiliation{\tunl} 
\author{M.~Busch}\affiliation{\duke}\affiliation{\tunl}	
\author{M.~Buuck}\altaffiliation{Present address: SLAC National Accelerator Laboratory, Menlo Park, CA 94025, USA}\affiliation{\uw} 
\author{T.S.~Caldwell}\affiliation{\unc}\affiliation{\tunl}	
\author{Y-D.~Chan}\affiliation{\lbnl}
\author{C.D.~Christofferson}\affiliation{\sdsmt} 
\author{P.-H.~Chu}\affiliation{\lanl} 
\author{M.L.~Clark}\affiliation{\unc}\affiliation{\tunl} 
\author{C.~Cuesta}\affiliation{\ciemat}	
\author{J.A.~Detwiler}\affiliation{\uw}	
\author{Yu.~Efremenko}\affiliation{\ut}\affiliation{\ornl}
\author{H.~Ejiri}\affiliation{\ou}
\author{S.R.~Elliott}\affiliation{\lanl}
\author{G.K.~Giovanetti}\affiliation{\williams}  
\author{M.P.~Green}\affiliation{\ncsu}\affiliation{\tunl}\affiliation{\ornl}   
\author{J.~Gruszko}\affiliation{\unc}\affiliation{\tunl} 
\author{I.S.~Guinn}\affiliation{\ornl}
\author{V.E.~Guiseppe}\affiliation{\ornl}	
\author{C.R.~Haufe}\affiliation{\unc}\affiliation{\tunl}	
\author{R.~Henning}\affiliation{\unc}\affiliation{\tunl}
\author{D.~Hervas~Aguilar}\affiliation{\unc}\affiliation{\tunl} 
\author{E.W.~Hoppe}\affiliation{\pnnl}
\author{A.~Hostiuc}\affiliation{\uw} 
\author{M.F.~Kidd}\affiliation{\ttu}	
\author{I.~Kim}~\email{inwookkim.physics@gmail.com}\altaffiliation{Present address: Lawrence Livermore National Laboratory, Livermore, CA 94550, USA}\affiliation{\lanl}
\author{R.T.~Kouzes}\affiliation{\pnnl}
\author{T.E.~Lannen~V}\affiliation{\usc} 
\author{A.~Li}\affiliation{\unc}\affiliation{\tunl} 
\author{J.M. L\'opez-Casta\~no}\affiliation{\ornl}
\author{E.L.~Martin}\altaffiliation{Present address: Duke University, Durham, NC 27708, USA}\affiliation{\unc}\affiliation{\tunl}	
\author{R.D.~Martin}\affiliation{\queens}	
\author{R.~Massarczyk}\affiliation{\lanl}		
\author{S.J.~Meijer}\affiliation{\lanl}	
\author{T.K.~Oli}\affiliation{\usd}  
\author{L.S.~Paudel}\affiliation{\usd} 
\author{W.~Pettus}\affiliation{\iu}\affiliation{\iuceem}	
\author{A.W.P.~Poon}\affiliation{\lbnl}
\author{D.C.~Radford}\affiliation{\ornl}
\author{A.L.~Reine}\affiliation{\unc}\affiliation{\tunl}	
\author{K.~Rielage}\affiliation{\lanl}
\author{N.W.~Ruof}\affiliation{\uw}	
\author{D.C.~Schaper}\affiliation{\lanl} 
\author{D.~Tedeschi}\affiliation{\usc}		
\author{R.L.~Varner}\affiliation{\ornl}  
\author{S.~Vasilyev}\affiliation{\JINR}	
\author{J.F.~Wilkerson}\affiliation{\unc}\affiliation{\tunl}\affiliation{\ornl}    
\author{C.~Wiseman}~\email{wisecg@uw.edu}\affiliation{\uw}
\author{W.~Xu}\affiliation{\usd} 
\author{C.-H.~Yu}\affiliation{\ornl}
\author{B.X.~Zhu}\altaffiliation{Present address: Jet Propulsion Laboratory, California Institute of Technology, Pasadena, CA 91109, USA}\affiliation{\lanl} 

\collaboration{{\sc{Majorana}} Collaboration}
\noaffiliation


\newcommand{\ackfunding}{This material is based upon work supported by the U.S.~Department of Energy, Office of Science, Office of Nuclear Physics under contract / award numbers DE-AC02-05CH11231, DE-AC05-00OR22725, DE-AC05-76RL0130, DE-FG02-97ER41020, DE-FG02-97ER41033, DE-FG02-97ER41041, DE-SC0012612, DE-SC0014445, DE-SC0018060, DE-SC0022339, and LANLEM77/LANLEM78. We acknowledge support from the Particle Astrophysics Program and Nuclear Physics Program of the National Science Foundation through grant numbers MRI-0923142, PHY-1003399, PHY-1102292, PHY-1206314, PHY-1614611, PHY-1812409, PHY-1812356, PHY-2111140, and PHY-2209530. 
We gratefully acknowledge the support of the Laboratory Directed Research \& Development (LDRD) program at Lawrence Berkeley National Laboratory for this work. 
We gratefully acknowledge the support of the U.S.~Department of Energy through the Los Alamos National Laboratory LDRD Program and through the Pacific Northwest National Laboratory LDRD Program for this work.  
We gratefully acknowledge the support of the South Dakota Board of Regents Competitive Research Grant. 
We acknowledge the support of the Natural Sciences and Engineering Research Council of Canada, funding reference number SAPIN-2017-00023, and from the Canada Foundation for Innovation John R.~Evans Leaders Fund.  
This research used resources provided by the Oak Ridge Leadership Computing Facility at Oak Ridge National Laboratory and by the National Energy Research Scientific Computing Center at Lawrence Berkeley National Laboratory, a U.S.~Department of Energy Office of Science User Facility. 
We thank our hosts and colleagues at the Sanford Underground Research Facility for their support.}

\date{April 11, 2024} 

\begin{abstract}

    Charge conservation and the Pauli exclusion principle result from fundamental symmetries in the standard model of particle physics, and are typically taken as axiomatic. 
    High-precision tests for small violations of these symmetries could point to new physics. 
    Here we consider three models for violation of these processes, which would produce detectable ionization in the high-purity germanium detectors of the \MJD\ experiment. 
    Using a 37.5 kg-yr exposure, we report a lower limit on the electron mean lifetime, improving the previous best limit for the $e \rightarrow \nu_e \overline{\nu_e} \nu_e$ decay channel by more than an order of magnitude. 
    We also present searches for two types of violation of the Pauli exclusion principle, setting limits on the probability of an electron to be found in a symmetric quantum state.

\end{abstract}


\maketitle

\section{Introduction}\label{sec:intro}

  Searches for small violations of fundamental symmetries have driven modern experimental physics, from the discovery of parity non-conservation in $\beta$-decay~\cite{wu1957parity}, to tests demonstrating violations of Bell's inequality~\cite{freedman1972experimental, aspect1982experimental, bouwmeester1997experimental}.
  In this work we consider two well-validated principles of quantum mechanics, charge conservation and Pauli exclusion, which emerge from robust mathematical frameworks and are typically taken as axioms.
  Many models have been proposed which allow their violation by exotic mechanisms~\cite{ignatiev1987search, greenberg1987local, greenberg1991particles, greenberg2000theories, bernabei2009new, abgrall2016search, okun1978cnc, mohapatra1987cnc}, and point to signs of new physics.

  Large underground radiation detectors offer a unique environment to search for rare signals produced by such symmetry violations.
  The \MJD, a high-purity germanium (HPGe) array, has world-leading energy resolution and ultra-low levels of radioactive backgrounds, and in addition to its primary search for neutrinoless double beta decay~\cite{aalseth2018search, majorana2023final}, it has been used to search for bosonic dark matter~\cite{vorren2017, mjd_exotic2022}, fractionally charged particles~\cite{massarczyk2018}, trinucleon decay~\cite{alvis2019trinucleon}, and signatures of quantum wavefunction collapse~\cite{mjd_wfcollapse2022}.

  The \DEM\ consists of two separate modules of p-type point contact HPGe detectors, with 29.7~kg enriched in $^{76}$Ge, and collected an ultimate exposure of 65~kg-yr~\cite{majorana2023final}.
  From this primary data set, an exposure of 37.5~kg-yr of \enrge\ data was selected for analysis of the 1--100 keV low-energy range.
  To produce the final spectrum, a series of analysis cuts are applied which remove events from electronics noise and energy-degraded surface events, while retaining bulk events above 20 keV with 92\% efficiency~\cite{mjd_exotic2022}.
  The \enrge\ detectors achieved background rates of 0.01~counts/(keV~kg~d) from 20--40 keV and 0.06~counts/(keV~kg~d) at 5 keV through use of highly radiopure materials, the deep underground location and careful control of the surface exposure time.
  In the energy spectrum, we observe a nearly flat continuum consisting of Compton scatter events between 20--100 keV, with visible contributions from $^{3}$H, $^{55}$Fe, and $^{68}$Ge below 20 keV.

\begin{figure*}
    \centering
    \includegraphics[width=\textwidth]{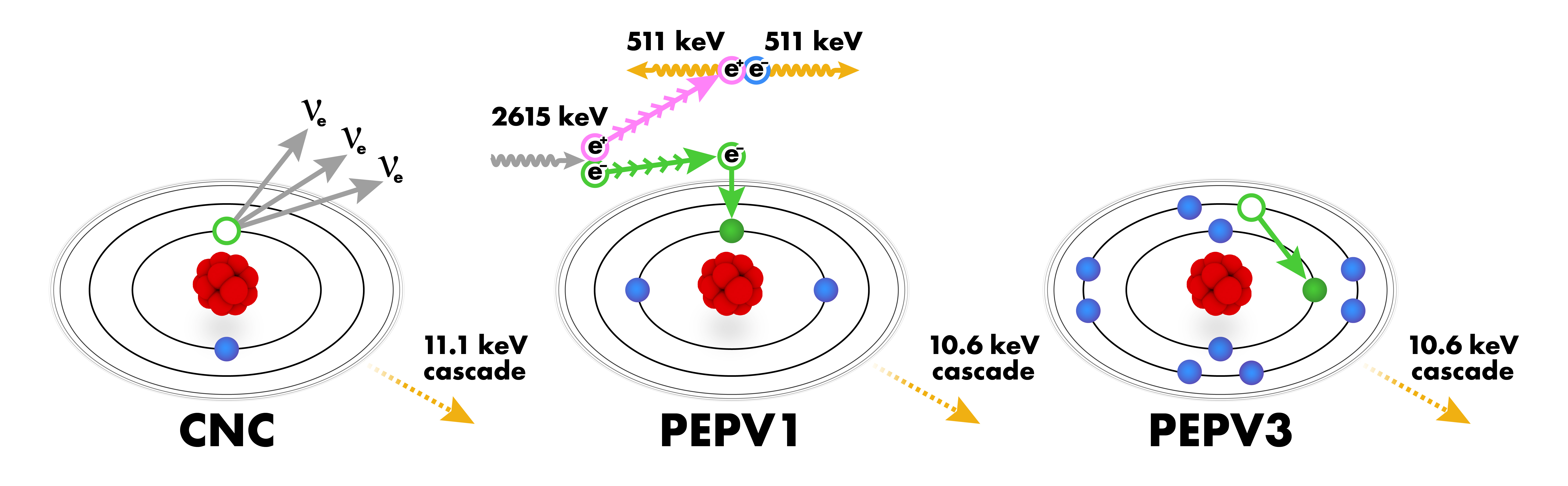}
    \caption{Three processes disallowed by quantum mechanics that would produce ionization in Ge atoms. 
    A simplified view of the atom includes a nucleus (red) surrounded by the closest orbital electrons (blue) depicting their nonstandard processes (green).  
    Additional bound electrons are not illustrated for clarity.
    In each case, electrons from outer shells cascade to fill the vacancy, releasing energy as photons and Auger electrons.
    \textit{Left:} Charge non-conservation, with an electron decaying to three neutrinos, releasing 11.1~keV.
    \textit{Middle:} Pauli exclusion principle (PEP) violation by a newly born electron produced by pair production from an incident 2615~keV gamma (Type I) releasing 10.6~keV along with two 511~keV gammas.  (Pink and green lines denote the positron/electron paths.)
    \textit{Right:} PEP violation where an electron descends to a fully occupied energy level (Type III), releasing 10.6~keV.
    }
    \label{fig:cartoon}
  \end{figure*}

  In this work we present an experimental test of charge conservation, searching for the spontaneous disappearance of an electron to ``invisibles'' (no photons), with the most favorable mode being to three neutrinos, $(e \rightarrow \nu_e \overline{\nu}_e \nu_e)$~\cite{workman2022_pdg}.
  Our result for the mean lifetime of the electron is the best in more than two decades~\cite{belli1999new}.
  We then report limits on violations of the Pauli exclusion principle, which would also have a detectable ionization signature in our HPGe array.
  The ``forbidden'' mechanisms considered are illustrated in Fig.~\ref{fig:cartoon}.

\section{Test of Electric Charge Conservation} \label{sec:cnc}

  Conservation of electric charge arises from the unbroken local U(1) symmetry of the Standard Model, with the photon as its associated massless boson.
  Extremely small experimental upper limits on the photon mass are generally considered to be evidence of exact electric charge conservation.
  However, there are theoretical frameworks beyond the Standard Model which allow electric charge non-conservation, either by broken gauge symmetry or by hidden processes such as charge leakage into extra dimensions~\cite{witten2018symmetry, chu1996cnc, ignatiev1979cnc, dubovsky2000electric}. 
  
  Violation of charge conservation implies that electrons, the lightest charged leptons, may have a finite lifetime.
  Hence, the conservation of electric charge can be tested by searching for the decay of electrons to chargeless particles with lighter mass, such as neutrinos and photons.
  Experiments have set limits on the decay process $(e \rightarrow \nu_e \gamma)$ by searching for a peak at 255.5~keV, with the best result from Borexino giving a mean lifetime $\tau_e > 6.6 \times 10^{28}$~y~\cite{borexino2015}.
  The electron may also decay without a photon to multiple neutrinos or other unknown chargeless beyond-Standard Model particles, often referred to as a ``disappearance'' mode.
  Decay to three neutrinos $(e \rightarrow \nu_e \overline{\nu}_e \nu_e)$ is considered the most favorable of these modes, being comprised of known particles which can balance angular momentum and conserve lepton number.
  In general, a search for electron disappearance would include effects from any ``invisible'' mechanism, not only the three-neutrino mode.
  Disappearance mechanisms to date give lifetimes on the order of $10^{24}$ years~\cite{workman2022_pdg}, and we point out that if more than one decay mode is available for an electron, the channel with the shorter lifetime will be favored.

  Data from the low-background physics run of the \DEM\ can be used to search for decay of atomic electrons within the HPGe detectors.
  If a K-shell (ground state) electron in a Ge atom decays to neutrinos (or other invisibles), a hole is produced in the shell, and electrons in higher shells will cascade to fill it, emitting X-rays and Auger electrons until the full binding energy~(11.1~keV) is released.
  This cascade occurs in a short time scale relative to the HPGe charge collection time, making the signature of this process a Gaussian peak in the spectrum at 11.1 keV.
  
  In the region of interest, we perform an unbinned profile likelihood scan over the number of counts attributable to a rare signal.
  The excellent energy resolution allows discrimination from the nearby 10.37 keV X-ray line originating from $^{68}$Ge electron capture decay, and the acceptance efficiency of the pulse shape analysis cuts is ($91\pm2$)\% at 11.1~keV.
  The energy calibration provided by the \nonubb\ analysis is validated by observation of the 10.37 keV line at the expected energy.
  The background function is a second-order Chebyshev polynomial, and nearby cosmogenic peaks are included in the model.
  This method is the same one used to search for peaked signatures in Ref.~\cite{mjd_exotic2022} and further details are given in the Methods section.

  \begin{figure}
    \centering
    \includegraphics[width=\columnwidth]{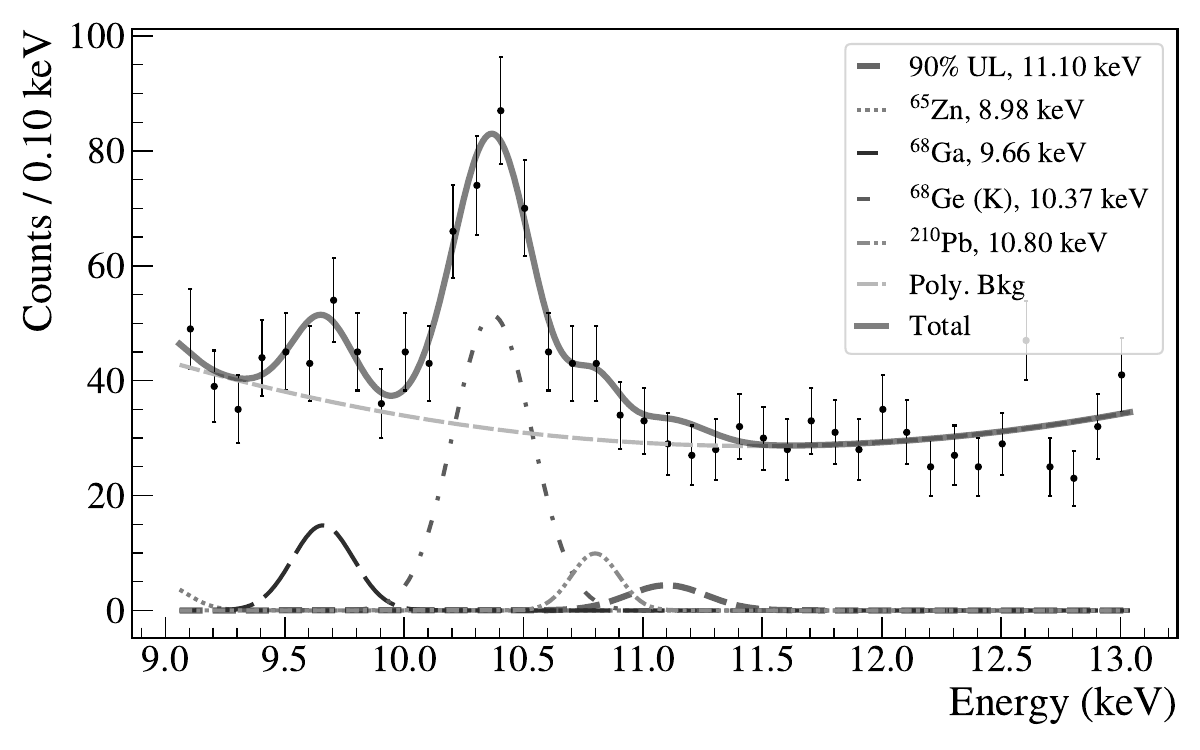}
    \caption{A charge nonconserving decay $e \rightarrow \nu_e \overline{\nu}_e \nu_e$ from a Ge K-shell electron would produce a peak at 11.1~keV.
    This peak is not observed in the 37.5~kg-y exposure collected, and an upper limit on the rate is set, $R=0.00154$~counts/kg-d (90\% CL).  Data error bars are Poisson-distributed, and the y-axis is given in counts per 0.1 keV bin.  Data is fit to a polynomial function with expected background lines from $^{65}$Zn, $^{68}$Ga, $^{68}$Ge, and $^{210}$Pb. Additional details are given in the Methods section.}
    \label{fig:cnc_fit}
  \end{figure}
  
  Finding no statistically significant signal at 11.1~keV, as shown in Fig.~\ref{fig:cnc_fit}, we report an upper limit on the event rate of $R = 0.00154$~counts/(kg-d) (90\% CL).
  The corresponding limit on the mean lifetime $\tau_e = n_\mathrm{e} / R$ is obtained from the upper limit on the decay rate per unit time $R$, the number density of Ge atoms, $N_\mathrm{Ge} = 7.96 \times 10^{24}$/kg, and two K-shell electrons for each Ge atom, $n_\mathrm{e} = 2N_\mathrm{Ge}$.
  We find a mean lifetime of $\tau_e(e \rightarrow \nu_e \overline{\nu}_e \nu_e) > 2.83 \times 10^{25}$~y (90\% CL), the most stringent limit for this decay channel by a factor 11.8 over Ref.~\cite{belli1999new}, surpassing the previous best result published more than two decades ago.

\section{Tests of Pauli Exclusion Principle Violation} \label{sec:pepv}

  The PEP states that two identical fermions cannot occupy the same quantum state~\cite{pauli1925}.
  In modern quantum mechanics, it is understood to originate from the spin-statistics theorem, which describes the antisymmetric behavior of fermions in quantum systems~\cite{weinberg1964feynman}.
  Many mechanisms of PEP violation have been proposed, making a direct comparison difficult~\cite{suzuki1993study, bernabei1997search, greenberg2000theories, bernabei2009new, Curceanu2017vip}.
  Experimental tests of the PEP may set limits on the probability of two fermions to form a symmetric quantum state.
  In this work, that probability is taken to be a ratio of lifetimes between PEP-obeying and PEP-violating atomic transitions of electrons, $\beta^2/2 \equiv \tau_\mathrm{PEP} / \tau_\mathrm{PEPV}$.

  In this case, the transitions in which a model allows the PEP to be violated are determined by the initial symmetry state of the electron.
  The Messiah-Greenberg superselection rule~\cite{messiah1964symmetrization} forbids electron transitions between states of differing symmetry, allowing only newly created electrons, or electrons already in a symmetric state, to make a transition to a symmetric final state.
  As electrons in symmetric states have not been observed, newly created ones provide the only model-independent test currently available.
  This constraint, however, can be evaded by exotic physics such as the existence of extra dimensions or electron substructure~\cite{greenberg1989phenomenology, akama1992superficial}.
  More recently, it has been proposed that the violation of the spin-statistics theorem and hence PEP violation can emerge naturally from quantum gravity~\cite{addazi2020modern}.
  The paper by Elliott et al.~\cite{elliott2012improved} reviews the experimental and theoretical considerations.
  Following this framework, processes that can violate the PEP are classified into three categories:
  \begin{itemize}
      \item \textbf{Type I} interactions are between a system of fermions and a fermion that has not previously interacted with any other fermions.
      For example, a newly created electron from pair production has not yet established (anti-)symmetry with the surrounding atomic lattice, and a PEP-violating process is allowed by any model.

      \item \textbf{Type II} interactions are between a system of fermions and a fermion that has not previously interacted with that given system.
      For example, an extant electron introduced to an atomic lattice (for example, through an electric current), may have new PEP-violating interactions with that lattice, despite having already established antisymmetry with respect to distant systems.

      \item \textbf{Type III} interactions are between a system of fermions and a fermion within that given system.  
      PEP violation in such interactions is only possible in models that avoid the Messiah-Greenberg superselection rule, since the PEP-violating fermion is already in an established symmetry state in its host system.

  \end{itemize}

  Each type of PEP violation can be tested experimentally with HPGe detectors, and tests of Types I and III are possible with the \DEM\ data set.
  In this work, a Type I search is performed using $^{228}$Th calibration data, and a Type III search is performed with the 37.5~kg-yr low-energy background data.
  Type II searches have been done previously by the \MJ\ and VIP collaborations, using electrical currents through Pb and Cu as the transition sources~\cite{elliott2012improved, piscicchia2020vip, napolitano2022testing}.
  We note that strong limits are available for both Type I and Type III processes based on searches for anomalous masses of primordial $^{5}$Li~\cite{thoma1992limits, nolte1991accelerator} and for forbidden nuclear transitions in $^{12}$C~\cite{borexino2010pep}. 
  However, both of these results consider transitions of strongly interacting nucleons in the potential generated by those same nucleons. 
  The searches performed here with the \DEM\ involve purely atomic transitions of electrons in a potential that is dominated by the electromagnetic attraction of the (positive) nucleus, and are the most stringent available in such systems. 
  As there is no comprehensive theoretical framework that accommodates PEP violation, we cannot directly compare our results here with the nuclear limits in Refs.~\cite{thoma1992limits} and~\cite{borexino2010pep}. 
  Instead, we stress that it is valuable to perform such complementary tests in as wide a variety of qualitatively different systems as possible.

\subsection{Type I PEP Violation test with new electrons} \label{sec:pepv1}

  The calibration system of the \DEM\ consisted of twin $^{228}$Th line sources, periodically inserted into the system for weekly 60-90 minute calibrations of each module, with several runs of longer duration for fine tuning of pulse shape analysis cuts.
  We utilize 40.43 days from the first detector module, and 21.38 days from the second module.
  The signature of Type I PEP violation in the \MJ\ calibration data set can be observed (Fig.~\ref{fig:cartoon}, middle scheme) from pair production events produced by 2614.5~keV gamma rays from the decay of $^{208}$Tl in the $^{228}$Th line source, which create electron-positron pairs in the detectors.
  The positron annihilation produces two 511~keV gamma rays, and one or both may escape the detector, creating single-escape peak (SEP) events at 2103.5~keV and double-escape peak (DEP) events at 1592.5~keV.
  If the PEP is violated, the pair-produced free electron may be captured by a Ge atom and transition to the already occupied K shell.
  In this process, the total binding energy of 10.6 keV is emitted, which is decreased from 11.1 keV since 3 electrons are present~\cite{elliott2012improved}.
  
\begin{figure}
    \centering
    \includegraphics[width=\columnwidth]{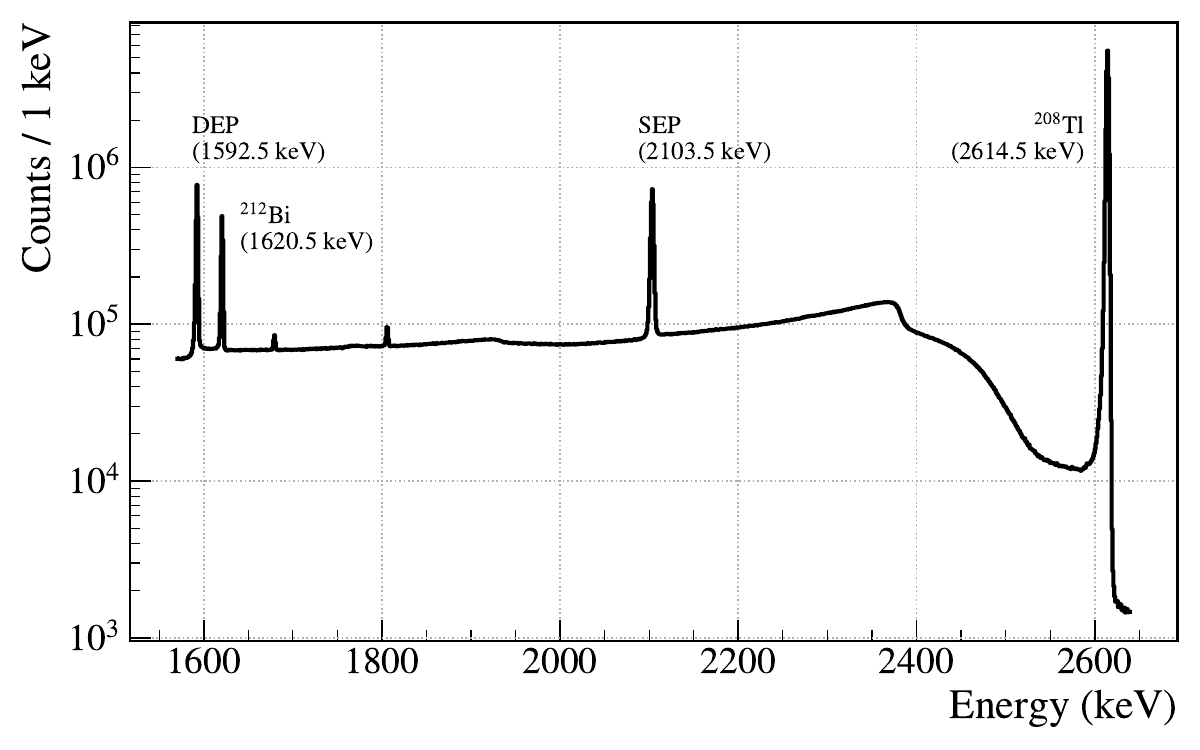}
    \caption{The combined $^{228}$Th calibration spectrum from all active detectors in the \MJD.  Prominent features include the full-energy peak from $^{208}$Tl, the associated single and double-escape peaks (SEP, DEP), and a strong $^{212}$Bi line near the DEP. 
    Data was taken from 2015--2019, presented here in counts per 1 keV bin.  Bin errors are Poisson-distributed.}
    \label{fig:FullSp}
  \end{figure}

  The full-energy peak at 2614.5~keV contains a significant contribution of ionization events with no pair production, which precludes its use in our search.
  The additional cascade produced by the PEP-violating capture to the K shell sums with the escape peak energy deposition, making the signature of the transition a peak 10.6~keV above the single-escape and double-escape peaks.
  The best prior limit for this process is $\beta^2/2 < 1.4\times 10^{-3}$ (99.7\% CL)~\cite{elliott2012improved} achieved by using a single HPGe detector and $^{232}$Th source, with three weeks of runtime.
  Our calibration runtime and active detector mass are both significantly larger, taken over the multi-year run of the \DEM.
  The spectrum used for our Type I search is shown in Fig.~\ref{fig:FullSp}, with standard data cleaning and quality cuts applied~\cite{alvis2019search}.

\begin{figure}
    \centering
    \includegraphics[width=\columnwidth]{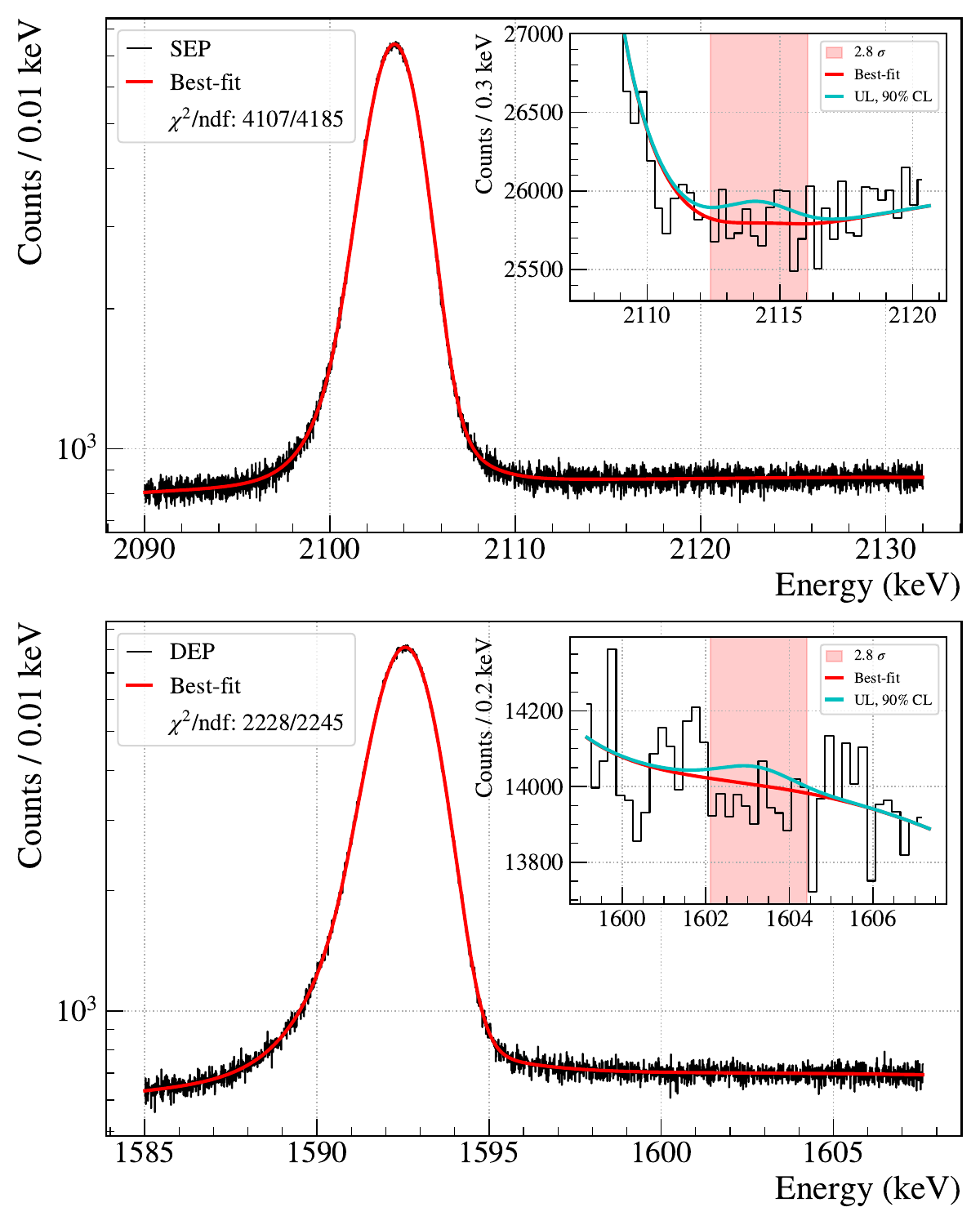}
    \caption{Searching for a violation of the Pauli exclusion principle via detection of an ``echo'' peak.
      Fits to the single-escape peak (SEP, top) and double-escape peak (DEP, bottom) regions are shown, with the best-fit (red) and the 90\% UL (cyan) in the inset.  A 2.8$\sigma$-wide region centered at the echo peak is shown for context.  Bin errors are Poisson-distributed, and the goodness-of-fit ($\chi^2/$ndf) of the model is close to 1 in both cases.  Additional details on the fits are given in the Methods section.}
    \label{fig:SEPDEPfits}
  \end{figure}

  To search the regions above the double-escape and single-escape peaks for a PEP-violating (``echo'') peak, we perform a standard extended binned likelihood fit, using the precision peakshape function given in Ref.~\cite{arnquist2023energy}.
  Fit results for both regions are shown in Fig.~\ref{fig:SEPDEPfits}.
  In addition to the main Gaussian term, it includes contributions modifying the high- and low-energy tails of the escape peaks, with a Legendre polynomial background centered in the fit window.
  These correction terms are essential, considering the large number of counts in the peaks.
  We treat the double-escape and single-escape peak regions independently, representing the ``echo'' peak by the same function, with its shape parameters determined by the immediately adjacent peak, and its energy fixed to 10.6~keV above the escape peak energy.
  The branching ratio $B$ determines the number of counts in the echo peak, relative to each escape peak.
  This approach improves over the assumption of a flat background made in Ref.~\cite{elliott2012improved}, where the upper limit was computed from a single bin with width 2.8 $\sigma$ in the echo region, which can bias the result if some curvature is present.

  In both escape regions, the data are consistent with a branching ratio of zero.
  In the SEP region a non-zero best-fit value is preferred, though its lower limit is still statistically consistent with zero signal ($< 1 \sigma$ significance).
  While the DEP region alone provides the most restrictive limit (at 90\% CL), the same physical process applies to both the double-escape and single-escape peaks, and we report the combined result as our primary limit, $\beta^2/2 < 3.69 \times 10^{-4}$ (90\% CL).
  This is currently the most stringent limit from an atomic Type I PEP violation search, improving the previous best limit by $\sim$70\%~\cite{elliott2012improved}. 

  Table~\ref{tab:results} gives exclusion limits for each region, and the peakshape parameters and profile likelihood scan results are given in Table~\ref{tab:bestfit} and Figure~\ref{fig:n2ll}.

  \begin{table}[h]
    \centering
    \caption{Upper limits on the $\beta^2/2$ PEP-violating parameter from analysis of the single-escape peak (SEP) and double-escape peak (DEP).
    While the double-escape peak result is more stringent (at 90\% CL), the same underlying mechanism applies to both escape peaks, and we report the combined result from both escape peaks as the primary result (bold text).}
    {\renewcommand{\arraystretch}{1.3}
    \begin{tabular}{c|cc}
      \hline \hline
        & \shortstack{$\beta^2/2$\\ 90\% CL} & \shortstack{$\beta^2/2$\\ 99.7\% CL} \\
      \hline
      DEP & $3.22 \times 10^{-4}$ & $8.32 \times 10^{-4}$ \\
      SEP & $9.99 \times 10^{-4}$ & $1.57 \times 10^{-3}$ \\
      \textbf{Combined} & \boldmath{$3.69 \times 10^{-4}$} & $7.79 \times 10^{-4}$ \\
      \hline \hline
    \end{tabular}
    }
    \label{tab:results}
  \end{table}

\subsection{Type III PEP Violation test with low-energy data}

  We also searched for Type III PEP-violating transition of an L-shell electron in a Ge atom to the already occupied K-shell~\cite{abgrall2016search, vorren2017}, shown in Fig.~\ref{fig:cartoon}.
  The signature of this process is a Gaussian peak at 10.6~keV, which would appear as a shoulder on the 10.37~keV $^{68}$Ge peak.
  Similar to the charge non-conservation (CNC) search, we search for the PEP-violating atomic transition with the 37.5~kg-yr low-energy data set.
  The total efficiency of the low-energy cuts is ($91\pm2$)\% at 10.6~keV, and we set an upper limit on the count rate in the spectrum at the region of interest with the same profile likelihood technique.

\begin{figure}
    \centering
    \includegraphics[width=\columnwidth]{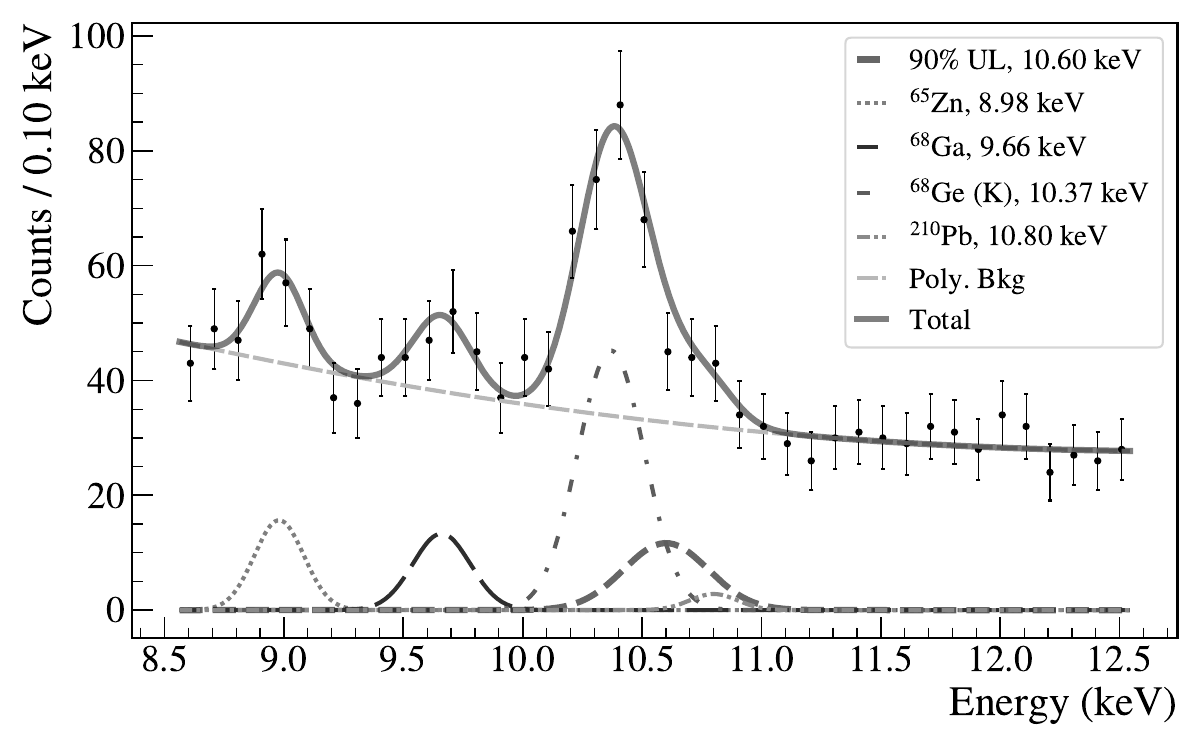}
    \caption{A PEP-violating transition of an L-shell electron to the fully-occupied K-shell in Ge would produce a peak at 10.6~keV.  
    This signature was not observed in the \DEM\ data set, and an upper limit on the rate is set, $R=0.0041$~counts/kg-d (90\% CL).  Error bars are Poisson-distributed, and the y-axis is given in counts per 0.1 keV bin.  
    As before, data is fit to a polynomial function with expected background lines from $^{65}$Zn, $^{68}$Ga, $^{68}$Ge, and $^{210}$Pb.}
    \label{fig:pepv_fit}
\end{figure}

  We find an upper limit on the count rate at 10.6~keV, $R = 0.0041$~counts/kg-d (90\% CL), shown in Fig.~\ref{fig:pepv_fit}.
  We then find the mean lifetime, $\tau_e = n_\mathrm{Ge} / R = 1.66 \times 10^{32}$~s (90 \% CL).
  \MJ\ previously set the most stringent upper limit at 90\% CL on this atomic Type III process, at $\beta^2/2 < 8.5 \times 10^{-48}$ with 478~kg-d exposure~\cite{vorren2017}.
  Comparing to the $1.7 \times 10^{-16}$~s mean lifetime of a standard K-$\alpha$ transition in Ge, we set an improved limit on the PEP-violating transition at $\beta^2/2 < 1.03 \times 10^{-48}$ (90\% CL), a factor 8.3 improvement over the previous limit~\cite{vorren2017}.

\section{Discussion}

  Low-background underground radiation detectors offer a unique enviroment to search for weak signatures from nonstandard processes.   
  The large datasets collected by \MJ\ allowed significant improvements in each search above existing limits; the CNC result is the best for ($e \rightarrow \mathrm{invisibles}$) since the 1999 DAMA result.
  This search based on atomic transitions is an important complement to nucleon-based searches such as Borexino~\cite{borexino2010pep}, as well as other atomic searches for PEP violation such as VIP-2~\cite{piscicchia2020vip, napolitano2022testing}.
  Larger Ge arrays with lower backgrounds are currently in active development~\cite{legendpcdr2021} and an order-of-magnitude improvement in these tests of fundamental quantum mechanical principles may soon be attained.

\begin{acknowledgments}

  This material is based upon work supported by the U.S.~Department of Energy, Office of Science, Office of Nuclear Physics under contract / award numbers DE-AC02-05CH11231, DE-AC05-00OR22725, DE-AC05-76RL0130, DE-FG02-97ER41020, DE-FG02-97ER41033, DE-FG02-97ER41041, DE-SC0012612, DE-SC0014445, DE-SC0018060, DE-SC0022339, and LANLEM77/LANLEM78.
  We acknowledge support from the Particle Astrophysics Program and Nuclear Physics Program of the National Science Foundation through grant numbers MRI-0923142, PHY-1003399, PHY-1102292, PHY-1206314, PHY-1614611, PHY-1812409, PHY-1812356, PHY-2111140, and PHY-2209530.
  We gratefully acknowledge the support of the Laboratory Directed Research \& Development (LDRD) program at Lawrence Berkeley National Laboratory for this work.
  We gratefully acknowledge the support of the U.S.~Department of Energy through the Los Alamos National Laboratory LDRD Program, the Oak Ridge National Laboratory LDRD Program, and the Pacific Northwest National Laboratory LDRD Program for this work.
  We gratefully acknowledge the support of the South Dakota Board of Regents Competitive Research Grant.
  We acknowledge the support of the Natural Sciences and Engineering Research Council of Canada, funding reference number SAPIN-2017-00023, and from the Canada Foundation for Innovation John R.~Evans Leaders Fund.
  We acknowledge support from the 2020/2021 L'Or\'eal-UNESCO for Women in Science Programme.
  This research used resources provided by the Oak Ridge Leadership Computing Facility at Oak Ridge National Laboratory and by the National Energy Research Scientific Computing Center, a U.S.~Department of Energy Office of Science User Facility.
  We thank our hosts and colleagues at the Sanford Underground Research Facility for their support.

\end{acknowledgments}

\bibliographystyle{apsrev4-1}
\bibliography{refs}

\appendix
\section{Methods}\label{sec:methods}
  
\subsection{Detectors and data taking scheme}

  The \DEM\ was located at the 4,850-ft underground level of the Sanford Underground Research Facility~(SURF) in Lead, South Dakota, consisting of two separate ultra-low background modules of p-type point contact HPGe detectors with a total mass of 44.1~kg, of which 29.7~kg were enriched to 88\% in $^{76}$Ge~\cite{abgrall2014majorana}. 
  The detectors were operated in a vacuum cryostat within a graded shield made from ultra-low background electroformed Cu and other shielding materials sourced to meet stringent material purity requirements~\cite{abgrall2016radioassay}.

  The \DEM\ calibration system accomodates a $^{228}$Th line source through a track penetrating the shield and surrounding each cryostat in a helical shape~\cite{mjdCalibration2017}.
  When deployed, the line source exposed all detectors in the array for energy calibration and stability determination. 
  During normal operations, sources were deployed weekly for 60-90 min to perform routine energy calibrations, while longer runs (12-24 h) were taken to refine energy and other pulse shape analysis parameters.

  From 2015--2019 the original set of 35 \enrge\ detectors were operated, using a data blindness scheme alternating 31 hours of open data followed by 93 hours of blinded data, to mitigate possible bias in the development of analysis routines.
  Analysis of the calibration data set does not employ a blinded approach.
  The \DEM\ continues to operate with 14.3~kg of natural Ge detectors in a single module for background studies and other rare-event searches.

\subsection{Peak fits at low energy}\label{sec:lowe_methods}

  The peak scanning algorithm uses a $\pm$(7$\sigma_\mathrm{res}$ + 1)~keV window near the peak position as the fit region of interest (ROI), where $\sigma_\mathrm{res}$ is the exposure-weighted combined detector resolution as a function of energy.
  The 1~keV offset ensures an ROI of at least 2~keV, even for a vanishingly small $\sigma$.
  The ROI for the PEP-violating transition search at 10.6~keV is 8.6--12.6~keV, and the ROI for the charge nonconserving electron decay at 11.1~keV is 9.1--13.2~keV.
  In the low-energy background data near the 10.6 keV signature, cosmogenic peaks at $^{65}$Zn (8.98~keV), $^{68}$Ga~(9.66~keV), and $^{68}$Ge~(10.37~keV) are expected.
  External $^{210}$Pb may also induce a peak at 10.8~keV.
  The energy calibration in the low-energy region is precise enough to treat the location of the cosmogenic peaks as fixed at the literature values.
  Other contributions from tritium beta decay and Compton scattering from higher energy peaks form a background continuum that is separable from the peak-like signature, and is approximated by a second-order polynomial within the ROIs.

  In the narrow ROIs where the background continuum can be approximated as a second-order polynomial, the spectrum is modeled as
  \begin{equation}
      \mathcal{P}(E; \vec{\theta}) = \eta(E) \Big( n_0 \mathcal{P}_\mathrm{poly} + \sum^{n_\mathrm{pks}}_i n_i\ \mathcal{P}_{G,i} +  n_\mathrm{obs}\mathcal{P}_\mathrm{rare}~\Big),
  \end{equation}

  \noindent
  where $\mathcal{P}_\mathrm{poly}, \mathcal{P}_{G,i}$ and $\mathcal{P}_\mathrm{rare}~$ are the normalized spectral distributions for the polynomial background as a function of energy $E$, the $i$-th cosmogenic peak, and the rare event peak of interest, respectively, and $n_0, n_i$ and $n_\mathrm{obs}$ are the number of events in each distribution.
  Additional nuisance parameters are denoted $\vec{\theta}$.
  We fit the model spectrum to the data using the unbinned extended likelihood method~\cite{wilks1938large, rolke2005limits, james2006statistical}.
  While the shape parameters for the polynomial background are unbounded, the widths of the Gaussian rare peak and background peaks are constrained by the detector energy resolution.
  Profiling the likelihood function by varying the number of counts in the rare signal peak and re-fitting the nuisance parameters at each step is the standard technique which results in a conservative upper limit on the rate.
  The signal peaks in both the CNC and Type III PEP searches overlap to some extent with the cosmogenic lines. 
  The likelihood contour is typically flat up to the number of counts assigned to the overlapping peaks, and only increases to the desired CL above these counts; hence the presence of overlapping peaks does not result in an artificially better limit.
  Fig.~\ref{fig:pepv_fit} shows the spectral fit in the ROI of the PEP-violating atomic transition in Ge at the 90\% CL upper limit.
  
  The total acceptance efficiency $\eta(E)$ of the low-energy data cleaning cuts is determined by convolving the data cleaning efficiency, the individual detector threshold efficiencies, and the efficiency of the energy-degraded slow pulse event rejection~\cite{wiseman2019}.
  The low-energy cuts retain $(91 \pm 2)$\% of single-site events in our ROI, which we take to be constant within each narrow fit window.
  We note that the efficiency correction is a smooth function and will not introduce peak structures, and remains above 80\% acceptance down to 3 keV~\cite{mjd_exotic2022}.
  Complementary searches for other PEP-violating atomic transitions, including jumps to the L-shell ($\sim$ 1.3 keV) and other between-shell jumps, are possible in principle.
  However, in the \DEM's energy spectrum, the L-shell peak is not resolvable from the background and in a region of steeply dropping acceptance efficiency close to the analysis energy lower limit.

\subsection{Peak fits at high energy}

  Our search for the PEP-violating ``echo'' peaks, located 10.6~keV above the SEP and DEP, utilizes a standard binned profile likelihood analysis, implemented with the \texttt{iminuit} toolkit~\cite{iminuit2020}.
  We perform a fit to our signal model using the \texttt{ExtendedBinnedNLL} cost function.
  The two peak regions are fit independently, since the parameters of each echo peak are strongly constrained by the adjacent large-amplitude escape peak.
  The number of counts in the echo peak is determined by the product of the number of counts in the escape peak times the branching ratio $B$, which is equal to the PEP-violating parameter $\beta^2/2$.
  If its best-fit value is not statistically significant (consistent with $B=0$), we report upper limits on $B$ by profiling over the likelihood function.

  A recent report from \MJ~\cite{arnquist2023energy} gives a detailed model of the composition of the germanium peak shape as a function of energy, including several features which are only observable at the large counting statistics available to the full \MJ\ calibration data set.
  The energy regions selected are chosen where the Compton background can be accurately modeled by a quadratic term, avoiding a nearby gamma peak from $^{212}$Bi at 1620 keV.
  The peakshape function is a sum of a Gaussian with several correction terms, given by
  \begin{equation}
    \mathcal{P}(E) = \mathcal{G}(E) + \mathcal{T}_\mathrm{HE}(E) + \mathcal{T}_\mathrm{LE}(E) + \mathcal{S}(E).
  \end{equation}
  \noindent
  The normalized Gaussian function includes the full number of counts $A$ attributable to the peak, the mean energy $\mu$ and width $\sigma$, and the fractional counts attributable to the low- and high-energy tails $f_l$ and $f_h$:
  \begin{equation}
    \mathcal{G}(E) = \frac{A(1-f_l-f_h)}{\sqrt{2\pi}\sigma} \exp \bigg(\frac{-(E-\mu)^2}{2\sigma^2} \bigg).
  \end{equation}
  
  Both the high- and low-energy tails are described by an exponentially modified Gaussian function
  \begin{align}
    \begin{split}
    \mathcal{T}_\alpha(E) &= \frac{A f_\alpha}{2 \gamma_\alpha} \exp \bigg(\frac{\sigma^2}{2\gamma_\alpha^2} \pm \frac{E-\mu}{\gamma_\alpha} \bigg) \\
    &\times \mathrm{erfc}\bigg(\frac{\sigma}{\sqrt{2}\gamma_\alpha} \pm \frac{E-\mu}{\sqrt{2}\sigma}\bigg).
    \end{split}
  \end{align}
  \noindent
  Here, the index $\alpha = \pm 1$ corresponds to the $\pm$ sign choice of high-(low-)energy tail, and $\gamma_l, \gamma_h$ the corresponding decay constant.
  A smoothed step function $\mathcal{S}(E)$ with fractional height $H$ is included to account for events which have undergone small-angle Compton scattering, and not contributing directly to the main peak.
  It is defined in terms of the complementary error function (erfc).
  It can be positive for incoming gammas, and negative for outgoing gammas.
  \begin{equation}
    \mathcal{S}(E) = \frac{A H}{2} \mathrm{erfc} \bigg(\frac{E-\mu}{\sqrt{2}\sigma}\bigg).
  \end{equation}
  Finally, the background function $\mathcal{B}(E)$ is modeled by a second-order sum of Legendre polynomials ($P_{1,2}$) centered in the fit region is used to describe the background in the escape and echo peak regions, using three free parameters, $a_0, a_1$, and $a_2$.
  The fit window is defined between low and high energies $E_\mathrm{lo}$ and $E_\mathrm{hi}$, with the center at $E_\mathrm{cen}$.
  \begin{equation}
    \mathcal{B}(E) = a_0 + a_1 P_1(E - E_\mathrm{cen}) + a_2 P_2(E - E_\mathrm{cen}).
  \end{equation}
  \noindent
  Centering the terms in the fit region ($E_\mathrm{cen} = (E_\mathrm{hi} - E_\mathrm{lo})/2)$ avoids correlations in the fit parameters and ensures a more stable fit for all values of the branching ratio $B$.
  
  The full signal model $\mathcal{M}$ consists of contributions from the two peaks and the background term, 
  \begin{align}
    \begin{split}
    \mathcal{M}(E) &= \mathcal{P} (E, A, \mu, \vec{\theta}) \\
     &+ \mathcal{P} (E, B \cdot A, \mu + 10.6 \mathrm{keV}, \vec{\theta}) \\
     &+ \mathcal{B} (E, a_0, a_1, a_2, E_{lo}, E_{hi}).
    \end{split}
  \end{align}
  \noindent
  The nuisance parameters common to both peaks are given by 
  \begin{equation}
    \vec{\theta} = (\sigma, f_l, f_h, \gamma_l, \gamma_h, H).
  \end{equation}

  To ensure a robust fit, initial guesses for $A, \mu, \sigma, H$, and so on are numerically estimated from the histogram, and a least-squares fit to the escape peak and background is performed.
  Using these results, the echo peak is included in a full likelihood fit to obtain the best-fit values for all parameters, including the branching ratio $B$.  
  Table~\ref{tab:bestfit} gives the best-fit results and corresponding uncertainties on all parameters from each fit.

  \begin{table}[h]
    \caption{Best-fit parameters for the germanium peak shape function in the single-escape peak (SEP) and double-escape peak (DEP) regions. 
    The branching ratio to the PEP-violating peak $B$ is included, as well as the amplitude (counts) parameter $A$, the mean energy $\mu$, and the shape parameters $\sigma, f_{l, h}, \gamma_{l, h}, H$, background polynomial coefficents $a_i$, and fit regions $E_{lo}, E_{hi}$.  
    Errors are calculated using the MINOS algorithm in \texttt{iminuit}~\cite{iminuit2020}.
    }
    {\renewcommand{\arraystretch}{1.5}
    \begin{tabular}{l|c|c}
      \hline \hline
       & DEP & SEP \\
      \hline
      $B$ & $(7.0 \times 10^{-13}) {}^{+1.3 \times 10^{-4}}_{-7.0 \times 10^{-13}}$ & $(3.0 \times 10^{-4}) {}^{+4.2 \times 10^{-4}}_{-3.0 \times 10^{-4}}$ \\
      $A$ & $1{,}779{,}000\ {}^{+6,000}_{-5,000}$ & $2{,}462{,}300\ {}^{+3,400}_{-3,400}$ \\
      $\mu$ & $1592.661\ {}^{+0.005}_{-0.005}$ & $2103.618\ {}^{+0.010}_{-0.009}$ \\
      $\sigma$ & $0.823\ {}^{+0.003}_{-0.003}$ & $1.298\ {}^{+0.005}_{-0.005}$ \\
      $f_{l}$ & $0.293\ {}^{+0.007}_{-0.007}$ & $0.302\ {}^{+0.012}_{-0.011}$ \\
      $f_{h}$ & $0.032\ {}^{+0.004}_{-0.004}$ & $0.075\ {}^{+0.005}_{-0.004}$ \\
      $\gamma_{l}$ & $1.050\ {}^{+0.040}_{-0.040}$ & $1.314\ {}^{+0.037}_{-0.035}$ \\
      $\gamma_{h}$ & $2.060\ {}^{+0.370}_{-0.310}$ & $1.640\ {}^{+0.100}_{-0.090}$ \\
      $H$ & $0.001\ {}^{+0.002}_{-0.002}$ & $0.001\ {}^{+0.001}_{-0.001}$ \\
      $a_0$ & $67{,}400\ {}^{+1{,}100}_{-1{,}200}$ & $84{,}300\ {}^{+400}_{-400}$ \\
      $a_1$ & $4{,}200\ {}^{+1{,}800}_{-1{,}800}$ & $3{,}900\ {}^{+700}_{-700}$ \\
      $a_2$ & $-2{,}200\ {}^{+800}_{-700}$ & $-1{,}400\ {}^{+400}_{-400}$ \\
      $E_{lo}$ & 1585.0 & 2090.0 \\
      $E_{hi}$ & 1608.0 & 2132.0 \\
      \hline \hline
      \end{tabular}
      }
    \label{tab:bestfit}
  \end{table}

\begin{figure}
    \centering
    \includegraphics[width=\columnwidth]{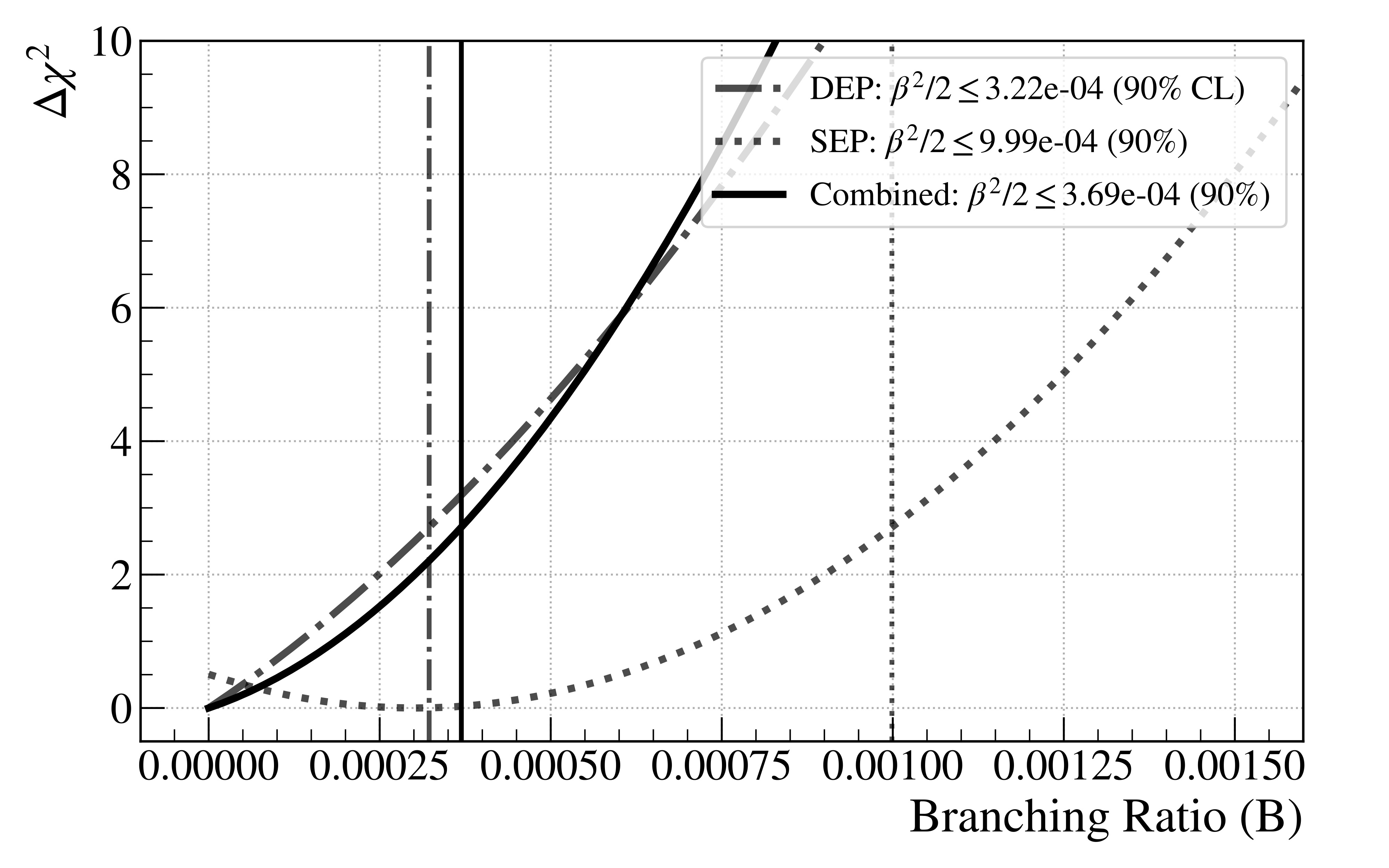}
    \caption{Profiles of the likelihood function over the branching ratio $B = \beta^2/2$, defined as the ratio of lifetimes between PEP-obeying and PEP-violating atomic transitions of electrons.
    The profiles from the single-escape peak region (dotted) and the double-escape peak region (dot-dashed) are added and shifted to produce the combined curve (solid black).  The upper limits are computed at the intersection $\Delta \chi^2 = 2.71$, representing the 90\% confidence level.  Vertical lines representing the 90\% upper limits are added for reference.}
    \label{fig:n2ll}
  \end{figure}

  Following the best-fit step, a profile of the likelihood function was run using the MINOS algorithm in \texttt{iminuit}, for the DEP and SEP search regions.
  Results from this scan are shown in Fig.~\ref{fig:n2ll}.
  While the best-fit for the DEP region is consistent with zero, the SEP region prefers a nonzero best-fit value.  
  Nonetheless, the lower limit in the SEP region does not exceed 1$\sigma$ significance, and we report the upper limit on $B$ in both cases.
  Adding the two likelihood curves, and computing the change in $\Delta \chi^2$ to the desired significance level, we see that the DEP result is the most restrictive at the 90\% CL, while the combined result is more restrictive at 99.7\% CL.
  However, since the same underlying physical process contributes to both regions equally, we quote the combined upper limit on $B$ (and hence $\beta^2/2$) as the primary result.

  Alternate analysis schemes exploiting the close-packed detector arrangement of the \DEM\ were also considered, including a technique of selecting only calibration events (multi-detector hits) where one detector records a 511 keV energy deposition, and searching for the same ``echo" peak signatures.
  For the \MJ\ data set, estimates indicate this method increases the signal-to-background (S/B) of the peaks by $\sim 15$\% but lowers the overall number of events available for analysis by 66\%.
  Near the SEP, we would require 1000 times more calibration data to make this method competitive, and the DEP requires another factor of 100 increase in statistics to provide similar results as the primary scheme which considers all available events.

\section{Data Availability}
  Source data containing the histograms in Figs.~\ref{fig:cnc_fit}, \ref{fig:FullSp}, \ref{fig:SEPDEPfits}, and \ref{fig:pepv_fit} are provided with this paper.

\section{Code availability}
  The analysis codes are available from the corresponding authors on reasonable request.

\section{Author Contributions}
  All authors were involved in various aspects of detector construction, operation, maintenance, data acquisition, data-taking shifts, software development, data processing and analysis.
  CW~and IK~developed the low-energy data cleaning, and low-energy statistical analysis and spectrum fit techniques.
  JMLC and CW conducted the high energy spectral analysis and derived the limit on the Type I PEP-violating process.
  IK~derived limits for the electron lifetime and Type III PEP-violating process.
  CW and IK wrote the manuscript.
  All authors reviewed, commented, and approved the results and the final version of the manuscript.

\end{document}